\documentclass[aip,reprint,amsmath,amssymb]{revtex4-1}

\usepackage{graphicx}
\usepackage{dcolumn}
\usepackage{float}
\usepackage{bm}
\usepackage{amssymb}

\begin{document}


\title{Polymer translocation into and out of an ellipsoidal cavity}

\author{James M. Polson}
\affiliation{%
Department of Physics, University of Prince Edward Island, 550 University Ave.,
Charlottetown, Prince Edward Island, C1A 4P3, Canada
}%

\date{\today}

\begin{abstract}
Monte Carlo simulations are used to study the translocation of a polymer into and out of a 
ellipsoidal cavity through a narrow pore. We measure the polymer free energy $F$ as a function
of a translocation coordinate, $s$, defined to be the number of bonds that have entered the 
cavity.  To study polymer insertion, we consider the case of a driving force acting on monomers
inside the pore, as well as monomer attraction to the cavity wall. We examine the changes to $F(s)$ 
upon variation in the shape anisometry and volume of the cavity, the polymer length, and the 
strength of the interactions driving the insertion. For athermal systems, the free energy 
functions are analyzed using a scaling approach, where we treat the confined portion of the 
polymer to be in the semi-dilute regime.  The free energy functions are used with 
the Fokker-Planck equation to measure mean translocation times, 
as well as translocation time distributions. We find that both polymer ejection and 
insertion is faster for ellipsoidal cavities than for spherical cavities. The results are
in qualitative agreement with those of a Langevin dynamics study in the case of ejection
but not for insertion. The discrepancy is likely due to out-of-equilibrium conformational
behaviour that is not accounted for in the FP approach.
\end{abstract}

\maketitle

\section{Introduction}
\label{sec:intro}

Polymer translocation is a fundamental process in which a polymer is transported 
through a narrow hole in a barrier divides two separate spaces.\cite{Muthukumar_book}
Over the past two decades, there has been considerable progress in the development of
techniques for detecting and monitoring single-molecule translocation events.
Much of this work has been motivated by the promise of an efficient and accurate
translocation-based method for nucleotide sequencing.\cite{%
carson2015challenges,steinbock2015emergence,wanunu2012nanopores,venkatesan2011nanpore}
Other technological applications include protein analysis,\cite{stefureac2008nanopore}
filtration of macromolecules,\cite{striemer2007charge} and controlled drug 
delivery.\cite{yang2012functionalized} Polymer translocation is also an important part of numerous biological
processes, including viral DNA packaging and ejection, transport of mRNA through
the nuclear pore complex, horizontal gene transfer between bacteria, and protein
transport across biomembranes.\cite{Alberts_book,Lodish_book} Due to its wide range 
of applications, polymer translocation has been the subject of numerous theoretical and 
computer simulation studies in recent years. Much of this work has been summarized in 
several recent reviews.\cite{Muthukumar_book,milchev2011single,panja2013through,palyulin2014polymer}

One important type of polymer translocation involves movement of polymers into or 
out of confined spaces.  Recent theoretical and computer simulation studies in this area 
have mainly focused on confinement in spherical or ellipsoidal cavities\cite{muthukumar2001translocation,%
muthukumar2003polymer,kong2004polymer,ali2004dynamics,ali2005coarse,cacciuto2006confinement,%
ali2006polymer,forrey2006langevin,ali2008ejection,sakaue2009dynamics,matsuyama2009packaging,%
ali2011influence,yang2012adsorption,rasmussen2012translocation,ghosal2012capstan,zhang2012dynamics,%
al2013effect,zhang2013dynamics,polson2013simulation,polson2013polymer,mahalik2013langevin,%
linna2014dynamics,zhang2014polymer,cao2014dynamics} or laterally unbounded spaces 
between flat walls.\cite{luo2009polymer,luo2010polymer,luo2011chain,sheng2012ejection}
Much of this work is motivated by the problems of viral DNA packaging and ejection,
in which DNA is confined to a space with dimensions comparable to that of its persistence 
length at near crystalline densities and very high internal pressures.
Experimental studies suggest that these processes do not follow simple quasistatic 
dynamics. For example, ejection proceeds in rapid transient bursts separated by 
pauses,\cite{chiaruttini2010vitro} while ultra-slow relaxation and nonequilibrium 
dynamics has been observed during packaging.\cite{Berndsen2014nonequilibrium}
A detailed understanding of these processes using theoretical methods requires,
at a minimum, the use of semi-flexible chain models to account for the high energetic
cost of confinement. Numerous studies have examined such models.\cite{ali2004dynamics,%
ali2005coarse,ali2006polymer,forrey2006langevin,%
ali2008ejection,ali2011influence,al2013effect,zhang2013dynamics,mahalik2013langevin,%
zhang2014polymer,cao2014dynamics} On the other hand, such studies are complemented
by those that use flexible-chain models to elucidate the specific role of conformational 
entropy on translocation.\cite{muthukumar2001translocation,muthukumar2003polymer,kong2004polymer,%
ali2004dynamics,cacciuto2006confinement,ali2006polymer,matsuyama2009packaging,%
yang2012adsorption,rasmussen2012translocation,zhang2012dynamics,zhang2013dynamics,%
polson2013polymer,linna2014dynamics} Other studies have examined the effects
of solvent quality,\cite{ali2008ejection}, electrostatic interactions,%
\cite{ali2011influence,cao2014dynamics}, temperature,\cite{al2013effect} and 
adsorption to the cavity surface.\cite{yang2012adsorption,rasmussen2012translocation} 

A few theoretical studies have considered the effect of cavity shape anisometry 
on polymer insertion and ejection.\cite{ali2006polymer,zhang2013dynamics,zhang2014polymer}
Ali {\it et al.} compared packaging and ejection in ellipsoidal and spherical cavities
of equal volume and found that flexible polymers package more quickly in spherical cavities
but eject faster in ellipsoidal cavities.\cite{ali2006polymer} By contrast, both processes 
are faster for spherical cavities in the case of semiflexible polymer, leading the
authors to suggest this as a reason for the spherical shapes of viruses with pressure-driven
ejection. Zhang and Luo recently used a 2-D system to study translocation of
a polymer into an elliptical cavity.\cite{zhang2013dynamics,zhang2014polymer}
For flexible chains, they found that the translocation time increased with cavity
anisometry.\cite{zhang2013dynamics} In the case of semiflexible polymers,
they found that the packaging rate depended the location of the entry point to
the cavity, as well as the packing fraction and chain stiffness.\cite{zhang2014polymer}
For a given elliptical cavity at high confinement, they found that entry along
the semi-minor axis gave the fastest packaging for sufficiently high 
stiffness.\cite{zhang2014polymer}

In many simulation studies of polymer translocation into or out of confined spaces, the 
scaling of the rate of transport with polymer length, $N$, and cavity dimension, $R$ have
been measured. Typically, the results are interpreted using estimates of the confinement free 
energy, $F$, of the polymer in the cavity and how it varies with the degree of translocation. 
The case of polymer ejection from spherical cavities provides a noteworthy illustration.
In an early study on the topic, Muthukumar assumed a scaling $F\sim N/R^{1/\nu}$ to describe
the confined portion of the chain, where $\nu\approx 0.588$ is the Flory scaling exponent. 
This was used to explain the observed scaling of the mean exit
time of $\tau\sim N (N/\phi)^{1/3\nu}$, where $\phi$ is the packing fraction of the
cavity for a fully inserted polymer.\cite{muthukumar2001translocation} 
Subsequently, Cacciuto and Luijten noted that the appropriate scaling of the free
energy for triaxial confinement is $F\sim N\phi^{1/(3\nu-1)}$. Using an approach 
suggested by the scaling of $\tau$ for driven translocation given in 
Ref.~\onlinecite{kantor2004anomalous}, they predicted $\tau \sim N^{1+\nu}\phi^{1/(3\nu-1)}$, 
which was consistent with Monte Carlo (MC) dynamics simulations.\cite{cacciuto2006confinement} 
Sakaue later questioned the implicit assumption of a constant free energy gradient 
during translocation and derived an alternative scaling relation that accounts for 
the decrease and eventual loss of entropic driving force near the end of the 
process.\cite{sakaue2009dynamics} The scalings observed in each of the MC dynamics studies
of Refs.~\onlinecite{muthukumar2001translocation} and \onlinecite{cacciuto2006confinement} 
were shown to be limiting cases of this more general result. 

An alternative approach to using analytical estimates of the translocation free energy
function is to calculate it explicitly for a chosen model using simulations. Rasmussen
{\it et al.} carried out such calculations for translocation of Lennard-Jones chains
into absorbing spherical cavities used the Incremental Gauge Cell MC method.%
\cite{rasmussen2012translocation} The free energy functions were then used in conjunction
with the Fokker-Planck (FP) equation to calculate translocation time distributions
and probabilities. They observed an interesting non-monotonic dependence of translocation 
times with a sharp peak located at a local free energy minimum.\cite{rasmussen2012translocation} 
These results were consistent with other results by the same group obtained using self-consistent 
field theory.\cite{yang2012adsorption} The validity of the FP approach requires quasistatic
conditions during translocation. As noted by Kantor and Kardar\cite{kantor2004anomalous} this
condition cannot be satisfied for long polymer chains. In addition, nonequilibrium behaviour
has been observed in simulations of polymer ejection.\cite{linna2014dynamics}
On the other hand, for sufficiently high pore friction quasistatic translocation
is possible and the FP approach is valid.\cite{Muthukumar_book} Recently, we used a
MC method to calculate translocation free energy functions\cite{polson2013simulation} 
and showed that the predicted translocation time distributions were perfectly
consistent with those obtained from dynamics simulations when the pore friction
was sufficiently high.\cite{polson2013polymer,polson2014evaluating}

The present work is a theoretical study of the translocation of a flexible polymer 
into and out of an ellipsoidal cavity. 
Following our other recent work,\cite{polson2013simulation,polson2013polymer,polson2014evaluating}
we use MC simulations to measure the translocation free energy functions. 
In addition to entropic effects, we also consider the effects of a force located 
in the nanopore that drives polymer
insertion into the cavity. We also consider the effect of monomer attraction to the
cavity wall, as in Ref.~\onlinecite{rasmussen2012translocation}. We examine the
effects on the free energy of varying several key system parameters, with special
attention to the effects of the anisometry of the cavity. The free energy functions
are used with the FP formalism to calculate translocation times for both polymer
insertion and ejection. We find that ejection is predicted to be
faster in ellipsoidal cavities in agreement with Ref.~\onlinecite{ali2006polymer}.
Interestingly, we also find that insertion tends to be slower for ellipsoidal cavities, in 
disagreement with results from that study. This points to the importance of nonequilibrium 
dynamics, which is not accounted for in the FP approach, as well as the limitations of 
using free energy functions to predict translocation dynamics.

\section{Model}
\label{sec:model}

We employ a minimal model of a polymer chain that translocates through a narrow pore between
an ellipsoidal cavity and semi-infinite space on one side of a flat wall. The polymer is 
modeled as a flexible chain of $N$ hard spheres, each with a diameter of $\sigma$. The pair potential for 
non-bonded monomers is thus $u_{\rm{nb}}(r)=\infty$ for $r\leq\sigma$, and $u_{\rm{nb}}(r)=0$ for 
$r>\sigma$, where $r$ is the distance between the centers of the monomers. Pairs of bonded monomers
interact with a potential $u_{\rm{b}}(r)= 0$ if $0.9\sigma<r<1.1\sigma$, and $u_{\rm{b}}(r)= \infty$,
otherwise.  Consequently, the bond length can fluctuate slightly about its average value.
Each monomer interacts with the walls of the system (pore, cavity and barrier) with a hard-particle
potential. Thus, the monomer-wall interaction potential is $u_{\rm w}(r)=0$ if the distance $r$ between 
the center of the monomer and the nearest point on the wall satisfies $r>0.5\sigma$, and 
$u_{\rm wall}=\infty$ if $r<0.5\sigma$. In addition, we consider two different potentials 
designed to drive the polymer into the cavity.  In the first case, monomers that lie inside 
the pore are subject to a potential described by a constant driving force of magnitude $f_{\rm d}$ 
that is directed toward the cavity.  In the second case, monomers inside the cavity whose centers 
lie a distance less than $\sigma$ from the wall have a potential energy of $-\epsilon$ 
(where $\epsilon>0$), while all other monomers have zero potential energy.

\begin{figure}[!ht]
\begin{center}
\vspace*{0.2in}
\includegraphics[width=0.38\textwidth]{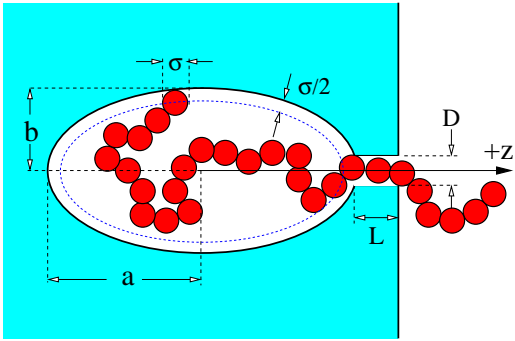}
\end{center}
\caption{Illustration of the system studied in this work. The polymer translocates
through a cylindrical pore of length $L$ and diameter $D$ between an ellipsoidal
cavity of semi-axis lengths of $a$ along $z$ and $b$ along $x$ and $y$.
The near-ellipsoidal space accessible to the monomer centers is shown bounded
by the dashed blue line. }
\label{fig:illust}
\end{figure}

The pore connecting the cavity to the open space is cylindrical in shape with a diameter $D$ and 
length $L$. For most simulations in this study, we choose $D=1.2\sigma$ and $L=1.3\sigma$.
We choose the $z$ axis to pass through the center of the cylindrical pore.
The semi-infinite space on one side of the pore is bounded by an infinite flat wall
perpendicular to $z$.
The cavity is an ellipsoid of revolution, with a semi-axis length of $a$ along $z$ and $b$
along $x$ and $y$. We consider the cases of both prolate ($a>b$) and oblate ($a<b$)
ellipsoids, as well as the special case of a spherical cavity ($a=b$). The volume $V$
of the ellipsoid is defined to be the volume of the subspace accessible to the centers of
the monomers. This subspace is enclosed by a virtual surface, each point on which lies
a distance of $0.5\sigma$ from the nearest point on the cavity wall. This virtual surface
deviates somewhat from ellipsoidal geometry, except in the special case where $a=b$.
The aspect ratio, $r$, of the cavity is defined to be the ratio of the dimensions
of the subspace, i.e. $r=(a-0.5\sigma)/(b-0.5\sigma)$.
The system is illustrated in Fig.~\ref{fig:illust}.

The degree to which the polymer has translocated across the nanopore into the cavity
is quantified using a translocation coordinate, $s$, which is defined in a manner 
similar to that in a recent study of ours.\cite{polson2014evaluating} It is equal to
the number of bonds that have crossed the mid-point of the nanopore. Typically,
one bond spans this point for any given configuration, and this bond contributes to $s$ 
the fraction that lies on the cavity side of the point. This is determined by the $z$ 
coordinates of the monomers connected by this bond.  Note that $s$ is a continuous 
variable in the range $s\in[0,N-1]$. 

\section{Fokker-Planck Formalism}
\label{sec:theory}

We define ${\cal W}(s,t;s_0,0)$ as the probability that a translocating polymer has 
coordinate $s$ at time $t$ given that it started at $s_0$ at time $t=0$.
In the quasi-static limit, ${\cal W}(s,t;s_0,0)$ is governed by the Fokker-Planck (FP) equation.  
In the case where the pore friction is sufficiently strong the equation has the 
form:\cite{Muthukumar_book}
\begin{eqnarray}
\frac{\partial{\cal W}(s,t;s_0,0)}{\partial t} & = & -\frac{\partial J(s,t;s_0,0)}{\partial s}
\label{eq:FP}
\end{eqnarray}
where the probability flux, $J(s,t;s_0,0)$ is defined
\begin{eqnarray}
J(s,t;s_0,0) = -{\cal D}\left[ \left(\frac{1}{k_{\rm B}T} \frac{\partial F}{\partial s}
{\cal W}\right) + \frac{\partial{\cal W}}{\partial s} \right],
\label{eq:Jdef}
\end{eqnarray}
where $k_{\rm B}$ is Boltzmann's constant and $T$ is temperature. Note that
the translocation rate constant ${\cal D}$ is independent of $s$ in this limit.

Consider a polymer with a translocation coordinate $s_0$ at time $t=0$, where
$s_0$ lies in the domain $s_0\in[s_{\rm a},s_{\rm b}]$. For $t>0$, the polymer
undergoes stochastic motion governed by ${\cal D}$ and $F(s)$ and eventually reaches
the domain boundary at $s=s_a$ or $s=s_b$. The first passage time,
$\tau$, is the time taken to reach $s=s_b$ for the first time without 
ever reaching $s=s_a$. The distribution of first passage times is given by 
the probability flux at $s=s_b$, i.e.
\begin{eqnarray}
P(\tau) = J(s_b,\tau;s_0,0),
\label{eq:Pdef}
\end{eqnarray}
where ${\cal W}$ and $J$ are calculated using adsorbing boundary conditions,
i.e. ${\cal W}(s_a,t;s_0,0)={\cal W}(s_b,t;s_0,0)=0$, and where the initial
condition implies ${\cal W}=\delta(s-s_0)$ at $t=0$.
This represents the probability per unit time that the polymer reaches 
the boundary at $s=s_b$. The probability that the system reaches $s_b$
first at any time is given by 
\begin{eqnarray}
p_b = \int_{0}^{\infty} P(\tau) d\tau,
\label{eq:pdef}
\end{eqnarray}
and the normalized probability distribution is given by 
\begin{eqnarray}
g(\tau;s_0) = P(\tau) / p_b.
\label{eq:gs0tau}
\end{eqnarray}
Finally, the associated mean first passage time is given by
\begin{eqnarray}
\langle\tau\rangle = \int_0^\infty d\tau\,\tau g(\tau;s_0).
\label{eq:taudef}
\end{eqnarray}
Similar relations can be obtained to describe translocation to the other boundary.

In this study, we use free energy functions obtained from MC simulations to 
solve Eq.~(\ref{eq:FP}) and (\ref{eq:Jdef}) subject to absorbing boundaries
at $s=0$ and $s=N-1$.
This is used to calculate the mean first passage time distributions in Eqs.~(\ref{eq:Pdef})
and (\ref{eq:gs0tau}), as well as the mean first passage time using Eq.~(\ref{eq:taudef})
and the translocation probability using Eq.~(\ref{eq:pdef}).
We use numerical integration methods, as described in the following section.

\section{Methods}
\label{sec:methods}


Monte Carlo simulations employing the Metropolis algorithm and the self-consistent
histogram (SCH) method\cite{frenkel2002understanding} were used to calculate the free energy functions
for the polymer-nanopore model described in Section~\ref{sec:model}. The SCH method
provides an efficient means to calculate the equilibrium probability distribution
${\cal P}(s)$, and thus its corresponding free energy function, $F(s) = -k_{\rm B}T\ln {\cal P}(s)$.
We have previously used this procedure to measure free energy functions in other
simulation studies of polymer translocation\cite{polson2013simulation,polson2013polymer,polson2014evaluating}
as well in a study of polymer segregation under cylindrical confinement.\cite{polson2014polymer}

To implement the SCH method, we carry out many independent simulations, each of which employs a
unique ``window potential'' of a chosen functional form.  The form of this potential is given by:
\begin{eqnarray}
{W_i(s)}=\begin{cases} \infty, \hspace{8mm} s<s_i^{\rm min} \cr 0,
\hspace{1cm} s_i^{\rm min}<s<s_i^{\rm max} \cr \infty, \hspace{8mm} s>s_i^{\rm max} \cr
\end{cases}
\label{eq:winPot}
\end{eqnarray}
where $s_i^{\rm min}$ and $s_i^{\rm max}$ are the limits that define the range of $s$
for the $i$-th window.  Within each ``window'' of $s$, a probability distribution $p_i(s)$ is
calculated in the simulation. The window potential width,
$\Delta s \equiv s_i^{\rm max} - s_i^{\rm min}$, is chosen to be sufficiently small
that the variation in $F$ does not exceed a few $k_{\rm B}T$. Adjacent windows overlap,
and the SCH algorithm uses the $p_i(s)$ histograms to reconstruct the unbiased distribution,
${\cal P}(s)$. The details of the histogram reconstruction algorithm are given in 
Ref.~\onlinecite{frenkel2002understanding}.  A description for an application to a physical 
system comparable to that studied here is presented in Ref.~\onlinecite{polson2013simulation}.

Polymer configurations were generated carrying out single-monomer moves using a combination of 
translational displacements and crankshaft rotations.  The trial moves were accepted with a
probability $p_{\rm acc}=\min(1,e^{-\Delta E/k_{\rm B}T})$, where $\Delta E$ is the energy 
difference between the trial and current states.  Initial polymer configurations were generated 
such that $s$ was within the allowed range for a given window potential. Prior to data sampling, 
the system was equilibrated.  As an illustration, for a $N=121$ polymer chain, the system was 
equilibrated for typically $\sim 10^7$ MC cycles, following which a production run of $\sim 10^8$ 
MC cycles was carried out.  During each MC cycle a move for each monomer is attempted once, on 
average.

The windows are chosen to overlap with half of the adjacent window, such that $s^{\rm max}_{i} =
s^{\rm min}_{i+2}$.  The window width was typically $\Delta s = \sigma$. Thus, a calculation for 
$N=121$, where the translocation coordinate spans a range of $s\in[0,120]$, required separate 
simulations for 239 different window potentials.  For each simulation, individual probability 
histograms were constructed using the binning technique with 10 bins per histogram.

The free energy functions obtained from the MC simulations were used with
the procedure summarized in Sec.~\ref{sec:theory} to study the translocation
dynamics under the assumption that conformational quasi-equilibrium is maintained
during the process.
The translocation probability ${\cal W}(s,t;s_0,0)$ was determined by solving
Eq.~(\ref{eq:FP}) for a chosen value of $s_0$. Typically, we used $s_0 = L/(2\sigma)$,
at which point the first monomer is just on the verge of exiting the pore into
the cavity. The equation was solved using standard numerical methods with a ``spatial'' 
grid size of $\Delta s=0.01$ and a time increment of $\Delta t = 0.002{\cal D}^{-1}$.
The distribution $P(\tau)$ was calculated using Eq.~(\ref{eq:Pdef}), where a standard
five-point method was used to evaluate the derivatives in Eq.~(\ref{eq:Jdef}).
The translocation probability, $p_b$, and mean first passage time, $\langle\tau\rangle$,
were calculated by numerical integration of Eqs.~(\ref{eq:pdef}) and (\ref{eq:taudef}),
respectively, using Simpson's rule.

In the results presented below, quantities of length are measured in units of $\sigma$, 
energy in units of $k_{\rm B}T$, force in units of $k_{\rm B}T/\sigma$
and time in units of ${\cal D}^{-1}$.

\section{Results}
\label{sec:results}

\begin{figure}[!ht]
\begin{center}
\vspace*{0.2in}
\includegraphics[width=0.42\textwidth]{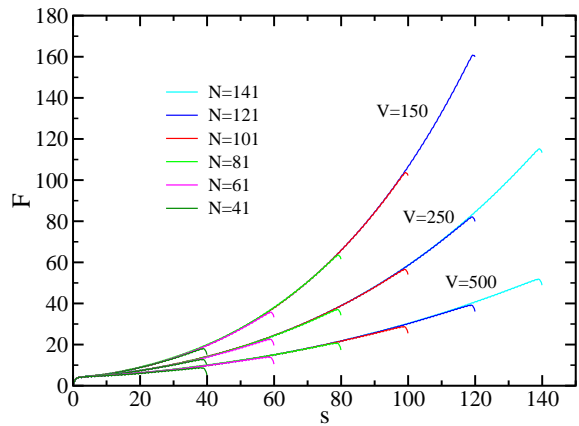}
\end{center}
\caption{Free energy functions for several different polymer lengths for
a cavity aspect ratio of $r=1$. Results for three different cavity volumes
are shown.  }
\label{fig:F.N.V.r1.0}
\end{figure}

We consider first the case of spherical cavities in the absence of a driving
force or adsorption potential, i.e. $r=1$, $f_{\rm d}=0$ and $\epsilon=0$. Figure~\ref{fig:F.N.V.r1.0} 
shows free energy functions for cavity volumes of $V$=150, 250 and 500, each for polymer 
lengths ranging from $N$=31 to $N$=141. As expected, the free energy cost of confining the
polymer in the cavity increases as the confinement volume decreases. In addition, the
curves all have positive curvature everywhere except at near the upper and lower bounds.
This indicates that free energy cost of inserting each monomer into the cavity
increases as the number (and hence density) of monomers inside increases.
This feature is consistent with results from previous MC studies\cite{rasmussen2012translocation,%
polson2013simulation,polson2013polymer} and theoretical studies\cite{kong2004polymer,yang2012adsorption}
that explicitly account for repulsion between monomers, in contrast to the case for ideal
polymers.\cite{muthukumar2003polymer} Another noteworthy feature is the strong degree of overlap 
between the curves for different $N$ and the same $V$. Thus, the free energy cost of transferring 
one monomer from the outside to the inside of the cavity depends approximately only on the density 
of monomers in the cavity.  Deviations from this trend are evident where $F$ dips abruptly 
near $s=N-1$. 

To provide a quantitative analysis of the free energy functions, we employ a standard 
scaling theory approach.\cite{sakaue2009dynamics} We first recall that the standard form of 
the entropic free energy barrier for translocation of a polymer through a narrow pore in a 
flat wall of negligible thickness is $F_0/kT = (1-\lambda)\ln[m(N-m)]$, where $m$ segments 
lie on one side of the pore, and $N-m$ lie on the other and where $\lambda=0.69$ is a critical 
exponent.\cite{Muthukumar_book} This expression can easily be modified to account for the case 
of a channel in a wall of finite thickness.\cite{polson2014evaluating} To account for the effect 
of confinement of the monomers inside the cavity, we consider the case where this part of the
polymer is in the semi-dilute regime. Here, the confined portion of the polymer can be 
viewed as a collection $n_{\rm b}$ blobs, each of size $\xi\sim \sigma\phi^{\nu/(1-3\nu)}$, 
where $\nu$ is the Flory exponent. 
It is easily shown that the confinement free energy scales as\cite{sakaue2009dynamics}
$\Delta F_{\rm c}/kT = n_{\rm b} \approx (V/\sigma^3)^{-1/(3\nu-1)} m^{3\nu/(3\nu-1)}$,
where $m$ is the number of monomers in the cavity. For a finite length nanochannel that
is spanned by an average of $n_{\rm p}$ bonds, we note that $m=s-n_{\rm p}/2$.
For the pore length $L=1.3$ used here, we estimate $n_{\rm p}=2$ when the pore is filled.
Thus, the total free energy, $F$, is expected to satisfy
\begin{eqnarray}
(F-F_0)/kT & \approx & (V/\sigma^3)^{-1.25} (s-n_{\rm p}/2)^{2.25}, 
\label{eq:FmF0}
\end{eqnarray}
where we have used $\nu=\frac{3}{5}$. This approximation is valid
when $n_{\rm p}/2<s<N-1-n_{\rm p}/2$. Over most of the range of $s$, $n_{\rm p}$ is
constant, the logarithmic term $F_0$ is negligible. Only near the upper and lower
bounds of $s$ is the variation of $F_0$ appreciable. This feature, as well as 
a partial emptying of the pore, accounts for the dips in $F$ near $s=N-1$.
Otherwise $F$ is dominated by the confinement free energy of the cavity.
In the case of fixed $V$, $F$ is predicted to increase with $s$ independent of the polymer
length $N$. In addition, the variation of $F$ with $s$ has positive curvature, and $F$
increases with decreasing $V$. These predictions are qualitatively consistent with the
data. 

The use of the semi-dilute approximation to estimate free energy of confinement in the cavity
is expected to be valid only over a restricted range of densities.  In Ref.~\onlinecite{cacciuto2006self} 
it was shown that the predictions are valid only up to packing fractions 
of $\phi\approx 0.15$, where $\phi\equiv \pi N\sigma^3/6V$. Beyond this value the number of monomers 
per blob is unacceptably low. A lower limit on $\phi$ is imposed by the condition that the 
number of blobs, $n_{\rm b}=N\phi^{1/(3\nu-1)}$, be sufficiently large. For the polymer 
lengths considered in this work, it is difficult to find a range of $s$ that satisfies 
both conditions simultaneously. To analyze the data, we use a more relaxed condition for
low density and consider the case where $n_{\rm b}\geq 3$. 

\begin{figure}[!ht]
\begin{center}
\vspace*{0.2in}
\includegraphics[width=0.42\textwidth]{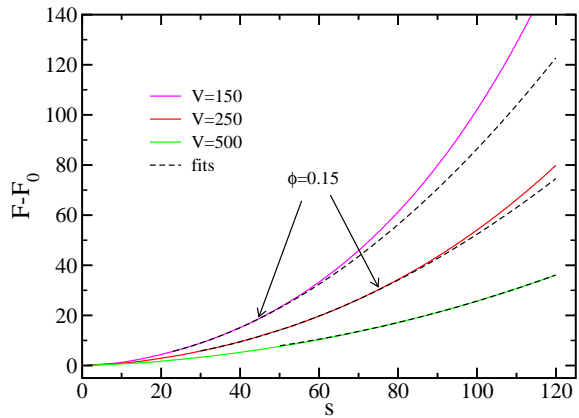}
\end{center}
\caption{Free energy functions for polymers of length $N=121$ and
a spherical cavity of various volumes. The dashed black lines show 
fits to the data in the region for which $n_{\rm b}\geq 3$
and $\phi<0.15$. The lower limit of this range is evident from the minimum
$s$ for the fitting curves, and the upper end of this range is explicitly
labeled for two of the curves. For $V$=500, the upper limit extends
beyond the range of the data.}
\label{fig:F.s.fit}
\end{figure}

Figure~\ref{fig:F.s.fit}
shows the results of fits using Eq.~(\ref{eq:FmF0}) for $N=121$ and various cavity
volumes. The free energy functions have been shifted by $F_0(s)$, which is the
free energy function for a flat wall that was calculated explicitly by simulation.
We use a fitting function of the form $F-F_0=c_0 + c_1(s-1)^{-\alpha}$.
The lower limit of the range of the fit is indicated in the plot, while
the fitting function is deliberately extrapolated beyond the upper limit of the
fitting range to illustrate the divergence of the prediction from the simulation
results at high density.  The upper limit itself is explicitly labeled for two of 
the functions. Note that scaling predictions underestimate the free energy in the 
region where $\phi\geq$0.15. This is consistent with the observation noted
in Ref.~\onlinecite{cacciuto2006self} that the confinement free energy crosses over 
into a concentrated regime where the excluded volume interactions are screened, which 
leads to a higher value of $\alpha$.  From the fit to the data in the valid range,
it was found that $\alpha$=1.9, 2.0 and 1.9 for $V$=150, 250 and 500, respectively. 
This is somewhat below the predicted value of $\alpha=2.25$. Noting the expected 
dependence of $c_1\propto V^{-\beta}$ where $\beta$=1.25, the ratio 
of $c_1$ measured for $V$=500 and $V$=250 yields $\beta=0.9$, while the
ratio for $V$=500 and $V$=150 yields $\beta$=1.1. Thus, the measured values
of the exponents $\alpha$ and $\beta$ both deviate from the predicted
values. Undoubtedly, this is arises from failing to properly satisfy the
condition that $n_{\rm b}\gg 1$.  We speculate that better satisfying this condition
would yield improved agreement. However, this would necessitate using polymer
chains at least an order of magnitude larger than those considered here,
which is not feasible for us at present. 

\begin{figure}[!ht]
\begin{center}
\vspace*{0.2in}
\includegraphics[width=0.42\textwidth]{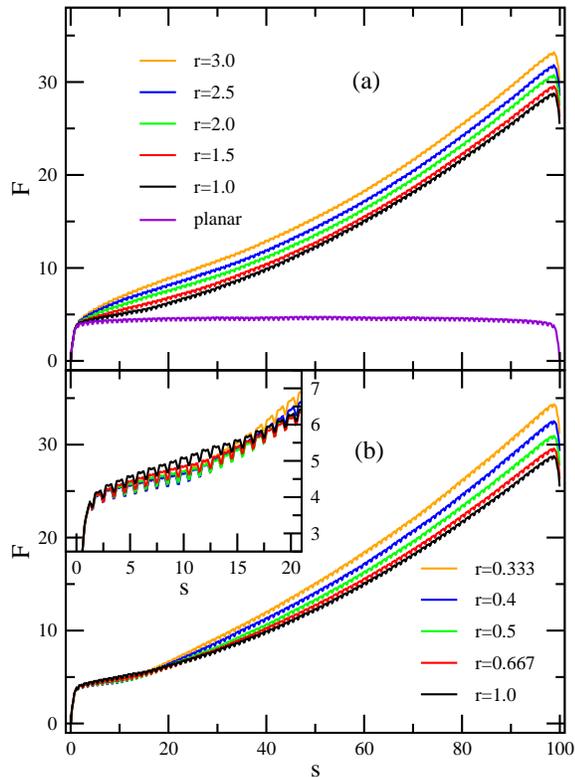}
\end{center}
\caption{Free energy functions for a polymer of length $N=101$ and 
a ellipsoidal cavity volume of $V=500$. No driving force is present.
Results for several different cavity aspect ratios are shown. The
graph in (a) shows results for prolate cavities and (b) shows results 
for oblate cavities. For comparison, the function $F_0$ for planar
wall geometry is shown in (a). The inset for (b) shows a close-up of 
the curves at low $s$.}
\label{fig:F.N101.V500}
\end{figure}

Next we consider the effects of the anisometry of the confining cavity.
Figure~\ref{fig:F.N101.V500} shows free energy functions for chains
of length $N=101$ in a cavity of volume $V=500$. Figure~\ref{fig:F.N101.V500}(a)
shows results for prolate ellipsoidal cavities ($r>1$)
and Fig.~\ref{fig:F.N101.V500}(b) shows results for oblate cavities ($r<1$).
In each case, the result for spherical cavities is also shown, for comparison.
The most notable feature here is the fact that deviations from spherical symmetry
in either direction lead to an increase in the free energy. Note, however, that
the curves for oblate cavities follow the opposite trend at low $s$ (see
the inset of Fig.~\ref{fig:F.N101.V500}(b)), in contrast to the case for 
prolate cavities.  The explanation for this difference is straightforward. 
As the first few monomers enter the cavity, the effects of confinement are felt
mainly by the curvature of the cavity wall near the pore. As the cavity becomes
more prolate, this local curvature increases, reducing the number of chain configurations,
and $F$ increases, accordingly. By contrast, as the cavity becomes increasingly
oblate, the local curvature decreases, leading to slight initial reduction of $F$
with decreasing $r$ evident in the figure. As more monomers enter the cavity
the polymer feels the presence of the cavity wall on the side opposite from the 
pore, and the trend reverses.

\begin{figure}[!ht]
\begin{center}
\vspace*{0.2in}
\includegraphics[width=0.44\textwidth]{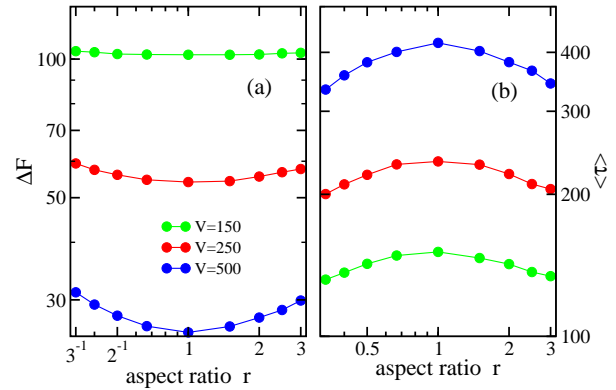}
\end{center}
\caption{(a) Free energy difference, $\Delta F\equiv F(N-1)-F(0)$, vs
ellipsoidal cavity aspect ratio, $r$. Results are shown for $N$=101
for three different cavity volumes. No driving force is present.
(b) Mean translocation time, $\langle\tau\rangle$, vs $r$.}
\label{fig:delF.tau.N101}
\end{figure}

Figure~\ref{fig:delF.tau.N101}(a) shows the variation of $\Delta F$
with the cavity aspect ratio $r$, where $\Delta F\equiv F(N-1)-F(0)$.
Results are shown for three different cavity volumes. For $V=500$,
$\Delta F$ is a minimum for spherical cavities, as was noted from
the results of Fig.~\ref{fig:F.N101.V500}. In addition, $\Delta F$
is approximately symmetrical about $r=1$ in the sense that
$\Delta F(r)\approx \Delta F(1/r)$, though oblate side is slightly
higher. For a lower cavity volume of $V=250$, the trend persists,
though with a weaker dependence of $\Delta F$ with $r$. For the lowest
cavity volume considered, $V=150$, there is negligible variation of
$\Delta F$ with $r$. Thus, the anisometry of the cavity affects the
confinement free energy in the cavity only at moderate densities,
while at high densities no effect is observable.

\begin{figure}[!ht]
\begin{center}
\vspace*{0.2in}
\includegraphics[width=0.42\textwidth]{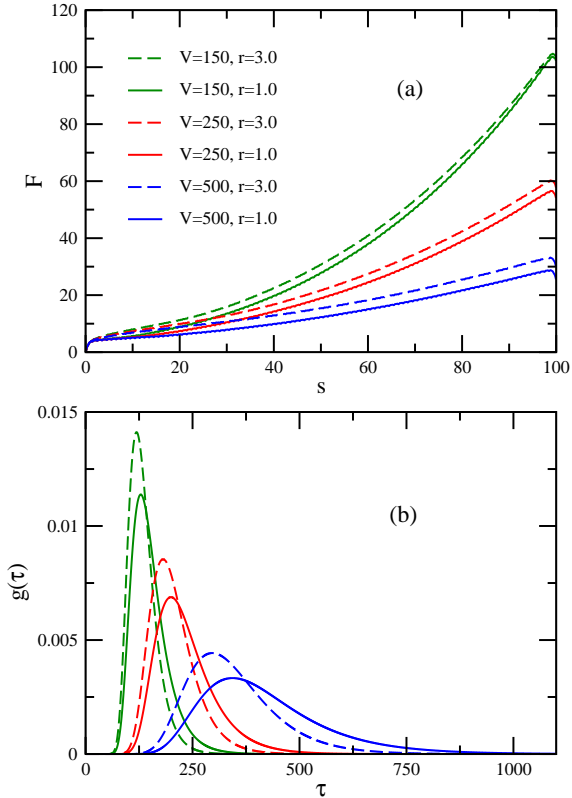}
\end{center}
\caption{(a) Free energy vs functions for a polymer of length $N=101$.
Data are shown for three different cavity volumes, each for spherical
($r=1$) and prolate ($r=3$) shapes. In all cases, the driving force
is zero. (b) Translocation time distributions calculated by numerically solving the FP
equation and employing the free energy functions shown in (a).
}
\label{fig:F.taudist.N101}
\end{figure}

Figure~\ref{fig:delF.tau.N101}(b) shows mean first passage translocation
times, $\langle\tau\rangle$, for ejection of the polymer from the cavity.
The results were obtained employing the FP formalism described earlier, using
the calculated free energy functions. As expected, $\langle\tau\rangle$
is a maximum for $r$=1 for $V$=500 and 250, and shows the same approximate
symmetry about $r$=1 as observed in (a). The dependence of $\langle\tau\rangle$
with $r$ does lessen slightly as $V$ decreases. Surprisingly, however, the
maximum of $\langle\tau\rangle$ at $r$=1 remains appreciable even at $V=150$,
in spite of the corresponding result for $\Delta F$. To understand this
apparent discrepancy, let us compare the free energy functions for $r$=1
and $r$=3 at each of the volumes considered.  The curves are shown in 
Fig.~\ref{fig:F.taudist.N101}. While the variation in $\Delta F$ with $r$ 
decreases with cavity volume, the free energy remains appreciably higher for 
the prolate cavity over most of the range of $s$. This corresponds to a somewhat 
higher average curvature of $F$ for the spherical cavity. Evidently, it is this
feature that leads to the observed difference in $\langle\tau\rangle$. 
Figure~\ref{fig:F.taudist.N101}(b) shows the full translocation time 
distributions obtained from the FP formalism, where the persistence of
the shift toward slower translocation for aspherical cavities for
smaller cavity volumes is clear. Note that the reduction of the ejection 
time with increasing cavity anisometry is consistent with the
results of the Langevin dynamics study of Ref.~\onlinecite{ali2006polymer}.

\begin{figure}[!ht]
\begin{center}
\vspace*{0.2in}
\includegraphics[width=0.42\textwidth]{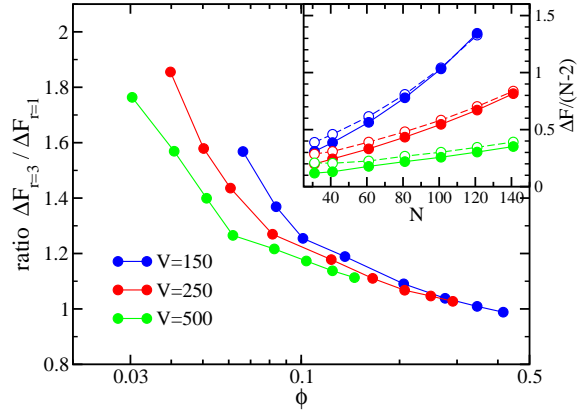}
\end{center}
\caption{Ratio of $\Delta F$ for $r=3$ and $r=1$ vs maximum packing fraction, $\phi$.
Data are shown for polymer lengths in the range $N$=31--141 and cavity volumes
of $V=$150, 250 and 500, each with aspect ratios of $r=1$ and 3. The inset shows the
scaled free energy barrier heights, $\Delta F/(N-2)$, vs polymer length, $N$. Solid
symbols correspond to $r=1$ and open symbols correspond to $r=3$.
}
\label{fig:delF.N.ratio}
\end{figure}

Next, we examine the effect of the variation of $\Delta F$ with the
cavity asymmetry upon changes in the polymer length. For a polymer at
$s$=$N-1$, there are $N-2$ bonds inside the cavity (one is still confined
in the pore). Thus, $\Delta F/(N-2)$ is the confinement free energy per
bond inside the cavity at this point.  The inset of Fig.~\ref{fig:delF.N.ratio} 
shows the $N$-dependence of $\Delta F/(N-2)$. Results are shown for three 
different cavity volumes, each with $r$=1 and $r$=3. Several trends are evident. 
First, for a cavity of fixed volume, the free energy per bond increases
with $N$, and it decreases with $V$. This follows from the fact that longer 
chains and smaller volumes correspond to higher packing densities inside 
the cavity. The other notable trend is that the confinement free energies 
for the prolate and spherical cavities are appreciably different at low
$N$ but tend to converge at high $N$. The convergence is stronger 
for smaller $V$. To clarify this trend, Fig.~\ref{fig:delF.N.ratio} shows
the ratio of $\Delta F$ for $r$=3 and $r$=1 plotted as a function of
$\phi$, the packing fraction in the cavity at $s=N-1$. Though the curves
for each $V$ do not overlap, they do follow the same overall pattern:
the ratio decreases monotonically toward unity as $\phi$ increases.
The same trend was found for oblate ($r=1/3$) cavities, as well (data not
shown). Thus, the anisometry of the cavity has negligible effect on the
confinement free energy at high density.

As noted in Ref.~\onlinecite{cacciuto2006self}, the confinement free energy of a 
polymer under triaxial confinement 
scales as $\Delta F_{\rm c} \sim \phi^{\nu/(1-3\nu)}$ in the semi-dilute 
regime, which crosses over to $\Delta F_{\rm c}\sim N\phi^2$ at higher density.
In neither case does the expression depend on the shape of the confining cavity.
However, these approximations likely require sufficiently large volumes to
apply. In the former case, the number of blobs is assumed to be large, implying
that the blob dimension $\xi$ is small relative to the confinement dimensions.
As $\phi$ decreases, $\xi$ increases, and the required condition becomes more
poorly satisfied. This, in part, may explain the trend observed in 
Fig.~\ref{fig:delF.N.ratio}. Clarification of these effects will require 
results from simulations employing much longer polymer chains.

Next, we consider the effect of a driving force acting on monomers inside
the pore to push the polymer into the cavity. These calculations
do not require additional simulations. To an excellent approximation, the
free energy function for finite $f_{\rm d}$ is given by
\begin{eqnarray}
F(s;f_{\rm d}) = F(s;0) - f_{\rm d}L s
\label{eq:Fsfd}
\end{eqnarray}
Thus, previous results for $f_{\rm d}=0$ can be easily modified to yield
free energy functions for finite $f_{\rm d}$.
A few simulations with a finite driving force were carried out, and the validity
of this expression was confirmed (data not shown). 

To study the effect of varying $V$ and $r$ on $F(s)$ and the translocation rate, 
it is helpful to introduce a modified driving force, defined
$f_{\rm d}^\prime\equiv f_{\rm d}-\Delta F(f_{\rm d}=0)/[L(N-1)]$.
Note that $f_{\rm d}^\prime=0$ corresponds to the case where $F(0)=F(N-1)$ for
arbitrary $f_{\rm d}$ and $V$.  The shift is designed to examine
the combined effect of the driving force and the {\it curvature} of $F(s)$
in the absence of the {\it entropic bias} toward polymer ejection.

\begin{figure}[!ht]
\begin{center}
\vspace*{0.2in}
\includegraphics[width=0.42\textwidth]{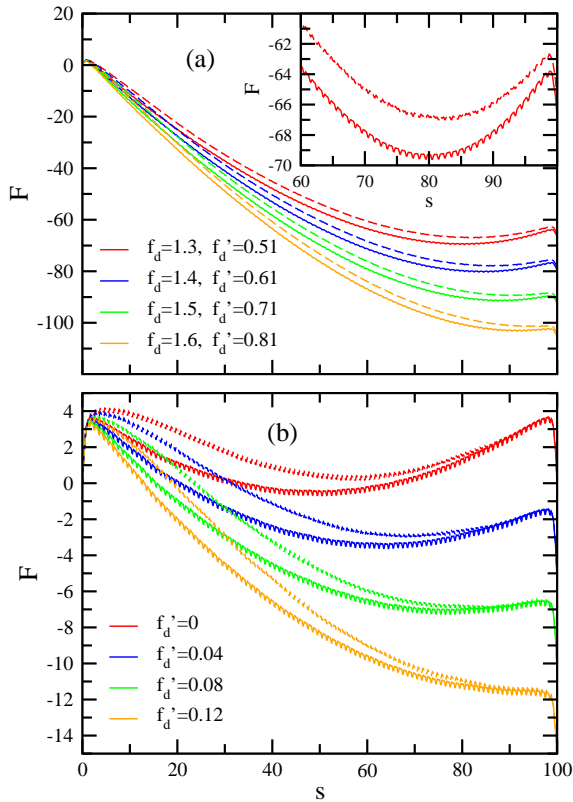}
\end{center}
\caption{Free energy functions for a polymer of length $N$=101 and 
a cavity of volume $V$=150 in (a) and $V$=500 in (b).  
Results are shown for several $f_{\rm d}$ and/or $f_{\rm d}^\prime$, 
where the shifted driving force $f_{\rm d}^\prime$ is defined in the text.
Solid curves correspond to spherical cavities (i.e. $r$=1) and
dashed curves are shown for prolate cavities with $r$=3. The inset
in (a) shows a close-up of the curves for $f_{\rm d}$=1.3.}
\label{fig:F.fd.N101.V150}
\end{figure}

Figure~\ref{fig:F.fd.N101.V150} shows results for polymers of length $N=101$
in a cavity of volume $V$=150 and $V$=500. Data are shown for several different driving
forces, each for the case of a spherical cavity and a prolate ($r$=3) cavity. 
The free energy curves in Fig.~\ref{fig:F.fd.N101.V150}(a)
are also labeled with $f_{\rm d}^\prime$. The curves in Fig.~\ref{fig:F.fd.N101.V150}(b)
for $V$=500 correspond to specific values of $f_{\rm d}^\prime$. For $V$=500,
note that $r$=1 and $r$=3 curves with the same $f_{\rm d}^\prime$ correspond
to different $f_{\rm d}$, since $\Delta F(f_{\rm d}=0)$ differ for different $r$,
as was evident in Fig.~\ref{fig:F.taudist.N101}(a).
The main effect of the driving force is to remove the free energy
penalty for inserting the polymer into the cavity. This point is first
reached when $f_{\rm d}^\prime$=0 (i.e. $f_{\rm d}=\Delta F(f_{\rm d}=0)/[L(N-1)]$.)
Beyond that point $F$ is lower for the polymer completely inside than where it is 
completely outside.  As shown for $V$=150 in Fig.~\ref{fig:F.fd.N101.V150}(a), increasing 
$f_{\rm d}$ generally increases rate that $F$ decreases with $s$. However,
the magnitude of $dF/ds$ decreases with $s$, which is a consequence of the 
curvature of $F(s)$ for $f_{\rm d}$=0. In addition, for moderate values of
$f_{\rm d}$, the function exhibits a local minimum when the polymer is mostly
inside the cavity. For example, for $f_{\rm d}$=1.3, the minimum lies
at $s_{\rm min}\approx$80, as is clear from the inset of the figure. The magnitude
of the resulting barrier between $s_{\rm min}$ and $s=N-1$ decreases 
and eventually vanishes about some critical value of $f_{\rm d}$.
The presence of such a minimum has been noted in previous studies of
translocation into spherical cavities with a driving force arising from attraction of
monomers to the cavity walls,\cite{yang2012adsorption,rasmussen2012translocation} 
and effective forces arising from translocation between cavities of different
sizes.\cite{kong2004polymer} 

As noted earlier, $\Delta F$ ($\equiv F(N-1)-F(0)$) is approximately invariant with respect to $r$
for $N$=101 and $V$=150 in the case where $f_{\rm d}$=0. From Eq.~(\ref{eq:Fsfd})
it follows that this is also true for arbitrary $f_{\rm d}$. However, over most
of the range of $s$, there is an appreciable dependence of $F(s)$ on $r$. This is
evident in Fig.~\ref{fig:F.fd.N101.V150}(a), which illustrates the difference in the
curves for the cases of $r$=1 and $r$=3. Unlike the case illustrated in 
Fig.~\ref{fig:F.taudist.N101} for $f_{\rm d}$=0, where the anisometry of the
cavity has a small effect on the rate for polymer ejection, small changes
in $r$ can have a significant effect on the rates of polymer insertion.

\begin{figure}[!ht]
\begin{center}
\vspace*{0.2in}
\includegraphics[width=0.42\textwidth]{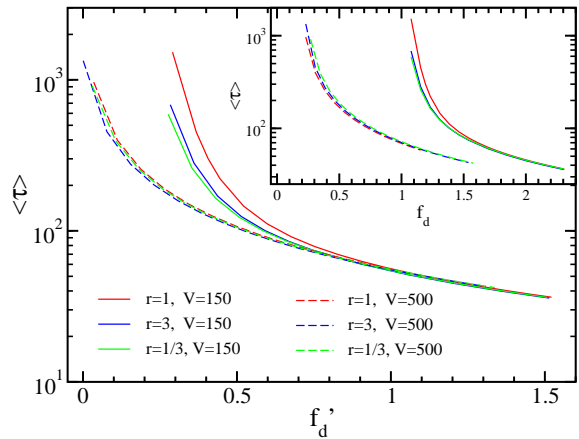}
\end{center}
\caption{Mean translocation time $\langle\tau\rangle$ vs shifted driving force, 
$f_{\rm d}^\prime$, where $f_{\rm d}^\prime \equiv f_{\rm d}-\Delta F(f_{\rm d}=0)/(L(N-1))$. 
Results are shown for a polymer of length $N$=101 for cavity volumes of $V$=150 
and $V$=500, each for three different cavity aspect ratios. The inset shows the 
same results plotted as a function of the driving force, $f_{\rm d}$.}
\label{fig:tau.fd.N101.scale}
\end{figure}

The effects of the free energy functions on the polymer insertion translocation times are
illustrated in Figure~\ref{fig:tau.fd.N101.scale}, which shows results for translocation times 
calculated using the FP method for $N$=101 and volumes of $V$=150 and $V$=500, each for 
cavity anisometries of $r$=1/3, 1, and 3. The inset shows $\langle\tau\rangle$ vs $f_{\rm d}$. 
As expected, the translocation times decrease with increasing $f_{\rm d}$. In addition, the
translocation rate is faster when the cavity volume is larger. This follows
from the fact that the entropic force acting against $f_{\rm d}$ is greater
for smaller volumes, as was noted in Fig.~\ref{fig:F.taudist.N101}(a).
The main part of the figure
shows $\langle\tau\rangle$ vs $f_{\rm d}'$.  At large $f_{\rm d}^\prime$, the translocation 
times all converge to the same curve. In this regime, effects of the curvature of $F(s)$
become negligible, and the translocation rate is governed principally by $dF/ds$.
For any given value of $f_{\rm d}^\prime$, $dF/ds\approx f_{\rm d}$
independent of $V$, and so $\langle\tau\rangle$ will be equal.
At lower values of $f_{\rm d}^\prime$, the results for the two volumes diverge. 
The curves for $V$=150 begin to increase rapidly near $f_{\rm d}^\prime\approx$0.5
and those for $V$=500 diverge near $f_{\rm d}^\prime\approx$0.1. In each case,
the rapid increase in $\langle\tau\rangle$ is due to the increasing influence
of the curvature of $F(s)$ as the driving force decreases. The curvature gives
rise to the free energy minimum evident in Fig.~\ref{fig:F.fd.N101.V150} and 
the associated free energy barrier. As was clear in Fig.~\ref{fig:F.taudist.N101}(a), 
the curvature of the function is greater when the cavity volume is lower. Thus, the 
onset of the rapid increase in $\langle\tau\rangle$ occurs first for $V$=150. 
For this volume, the free energy minimum first appears around $f_{\rm d}\approx 1.5$,
and deepens rapidly as $f_{\rm d}$ increases. As seen in Fig.~\ref{fig:F.fd.N101.V150},
this corresponds to $f_{\rm d}^\prime\approx 0.7$. By contrast, the free energy 
minimum for $V$=500 appears at $f_{\rm d}^\prime\approx$0.1. Both of these results 
are consistent with the trends in the figure.

Finally, we consider the effects of cavity anisometry on the translocation times.
At high $f_{\rm d}$, it is clear that varying $r$ has negligible effect on 
$\langle\tau\rangle$.  In this limit, $F(s)\approx -f_{\rm d}{L}s$, which is independent
of $r$, and so the invariance of $\langle\tau\rangle$ to $r$ is
expected. For $V$=150, a different trend emerges at lower $f_{\rm d}$, where 
the free energy minimum emerges. In this regime, $\langle\tau\rangle$ becomes 
greater for spherical cavities than for either prolate or oblate cavities.
For example, for $f_{\rm d}=1.15$, $\langle\tau\rangle$ is 1.6 times greater
for $r$=1 than for $r$=3. This effect is not present for $V$=500.
Thus, shape anisometry of the cavity leads to faster
polymer insertion at low driving force and high packing fraction.
This can be understood from inspection of Fig.~\ref{fig:F.fd.N101.V150}(a), where
we observe a deeper free energy minimum for $r$=1 than for $r$=3.  This trend persists
for functions with lower $f_{\rm d}$ than those values shown in figure. 
In the case of $V$=500, there appears to be negligible effect of the anisometry on 
$\langle\tau\rangle$. From the arguments above, there is expected to be some effect 
in the regime where a free energy minimum is present. From Fig.~\ref{fig:F.fd.N101.V150}(b),
this occurs for $f_{\rm d}^\prime \lesssim 0.1$ However, we note that the difference
in the depths of the minima for $r$=1 and $r$=3 is $\lesssim kT$. By contrast, 
the difference is considerably greater for $V$=150 at some positive values of $f_{\rm d}'$.
Consequently, the effects of the cavity anisometry are expected to be more appreciable
for smaller $V$, in accord with the results.

The translocation time predictions can be compared with the results of the Langevin
dynamics simulation study of Ali {\it et al.}\cite{ali2006polymer} They studied ejection
and insertion of flexible and semi-flexible polymers of length $N=100$ for a prolate
ellipsoidal cavity of comparable dimensions to the $V=150$ cavity considered here, as well 
as for a spherical cavity of the same volume.  In the case of ejection of a flexible
polymer, they found faster translocation for ellipsoidal cavities, in agreement
with the results presented earlier. However, for insertion of a flexible polymer, 
they found that translocation was faster for the spherical cavity. 
Those results are in disagreement with our predictions. Most likely, this is due
to out-of-equilibrium conformational behaviour in the dynamics simulations. Such 
effects are not accounted for in the FP approach, which assumes quasistatic dynamics.
The predictions are also in disagreement with the 2-D Langevin dynamics study in 
Ref.~\onlinecite{zhang2013dynamics}, where insertion proceeded most rapidly
in the case of a circular cavity, independent of packing fraction. In this case,
it is less clear whether the discrepancy arises from nonequilibrium dynamics or
from a different underlying behaviour of the free energy for 2-D. Regardless, these
results demonstrate the need to exercise caution in using free energy arguments
to interpret translocation simulation results. The assumptions for the valid
use of the FP approach are only satisfied for sufficiently high pore friction,
as we noted in Refs.~\onlinecite{polson2013polymer} and \onlinecite{polson2014evaluating}.
However, such predictions are useful for identifying some of the qualitative
effects of nonequilibrium dynamics.

\begin{figure}[!ht]
\begin{center}
\vspace*{0.2in}
\includegraphics[width=0.42\textwidth]{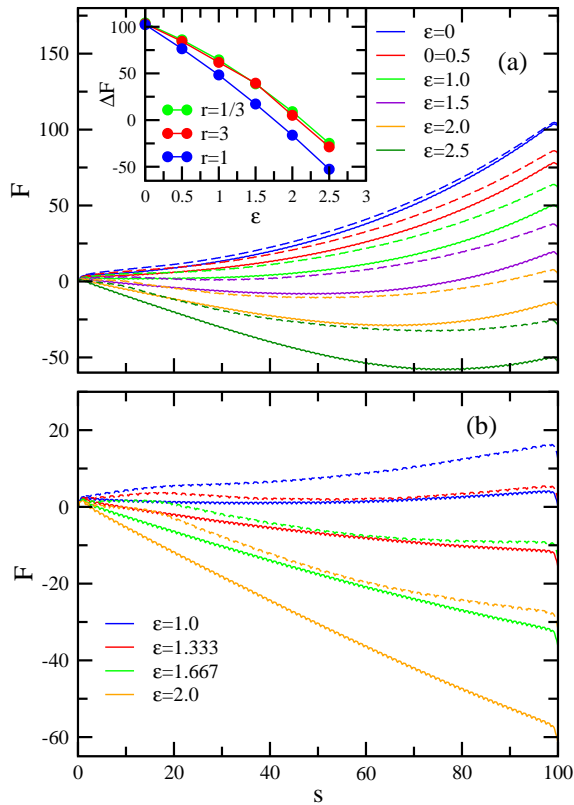}
\end{center}
\caption{Free energy functions for an adsorbing cavity for various adsorption strengths
for a polymer of length $N$=101. Results in (a) are for $V$=150 and (b) are for $V$=500.
The solid curves are for spherical ($r$=1) cavities, and the dashed curves are for
prolate ($r$=3) cavities. The inset in (a) shows the free energy difference 
$\Delta F\equiv F(N-1)$--$F(0)$ vs $\epsilon$.  }
\label{fig:F.eps.N101}
\end{figure}

The final case we consider in this study is translocation into an adsorbing cavity.
Here, monomers have an energy of $-\epsilon$ if their centers lie within a distance $\sigma$ (=1)
from the elliptical surface of the cavity. Figure~\ref{fig:F.eps.N101} shows free energy
functions for various values of $\epsilon$ for a polymer of length $N$=101. Results
are shown in (a) for a cavity volume of $V=150$ and in (b) for $V$=500, each for cavity 
anisometries of $r$=1 and 3.  As expected, increasing the attraction to the
cavity wall decreases $\Delta F$, and thus removes the entropic cost of insertion
of the polymer into the cavity. Generally, the free energy function has positive
curvature. Beyond some value of $\epsilon$, we see that $\Delta F<0$, i.e. it is more 
favourable for the polymer to lie completely inside rather than completely outside 
the cavity. However, there is an intermediate range of $\epsilon$ for which there
is a local free energy minimum and thus a free energy barrier that must be overcome 
for complete insertion. The depth of the free energy minimum is greater in the
cavity with the smaller volume.
These  trends are qualitatively comparable to those for the curves shown in 
Fig.~\ref{fig:F.fd.N101.V150} for the case of a driving force in the nanopore. 
In addition, the curves are qualitatively similar to those calculated in 
Refs.~\onlinecite{yang2012adsorption} and \onlinecite{rasmussen2012translocation}
for adsorbing spherical cavities. Note that the details of the model employed
in those studies differed from the present model, and thus a direct quantitative
comparison of the results is not possible.

For each $\epsilon$ value, the free energy curve for the prolate cavity lies above that 
for the spherical cavity. The results for oblate cavities with $r$=1/3 were comparable 
to those of $r$=3 (data not shown).  Thus, increasing the anisometry of the cavity reduces the 
effectiveness of the attraction to the cavity wall to drive the polymer into the cavity.
This general result is also clear in the inset of Fig.~\ref{fig:F.eps.N101}(a), which shows 
$\Delta F$ vs $\epsilon$. Comparing with Fig.~\ref{fig:F.fd.N101.V150}, we note that the 
cavity anisometry has the same effect on the free energy for the case of attractive cavity 
walls as it did for the case of a driving force acting in the pore. However, the differences
between the free energies for spherical and anisometric cavities are greater in the
present case, and increase with increasing $\epsilon$. Comparing the results in 
Fig.~\ref{fig:F.eps.N101}(a) and (b), we find that the absolute degree to which the
the curves are shifted between $r$=1 and $r$=3 is not significantly affected by 
the cavity volume. 

\begin{figure}[!ht]
\begin{center}
\vspace*{0.2in}
\includegraphics[width=0.42\textwidth]{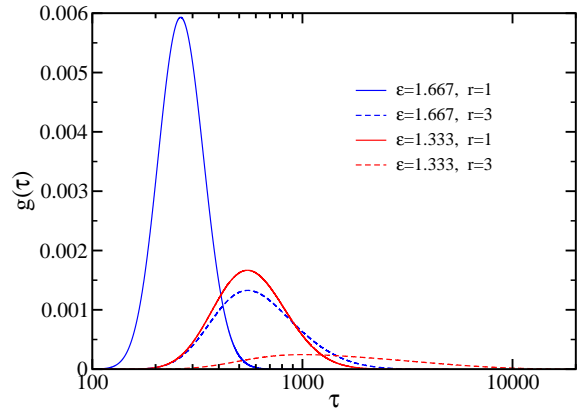}
\end{center}
\caption{Translocation time distributions for systems with an attractive cavity
wall, for $N$=101 and $V$=500. Results for two values of $\epsilon$ are shown,
each for cavity aspect ratios of $r$=1 and $r$=3.}
\label{fig:td.eps.N101.V500}
\end{figure}

Translocation time distributions calculated using the FP formalism and the free energy
functions in Fig.~\ref{fig:F.eps.N101}(b) are illustrated for $N$=101 and $V$=500 in 
Fig.~\ref{fig:td.eps.N101.V500}. Results are shown for values of $\epsilon$ that
are sufficiently large for an appreciable probability that the polymer is driven 
into the cavity. As expected from the free energy functions, increasing the cavity 
anisometry significantly slows down the rate of insertion. By contrast, cavity
anisotropy has a negligible effect on the polymer insertion rate for cavities
of the same volume in the case where insertion is driven by a force acting in
the nanopore, as was evident in Fig.~\ref{fig:tau.fd.N101.scale}. The strong
effect observed here apparently arises from the fact that changing the shape 
of the cavity distorts the polymer in a way that reduces the average number of 
monomer-surface contacts in the cavity for the accessible conformations.
For the model with the driving force in the nanopore, this kind of effect 
does not affect the energetic contribution to the free energy, unlike the case 
where surface interactions are present.

\section{Conclusions}
\label{sec:conclusions}

In this study, we have used simulation and theoretical methods to investigate
the translocation of a flexible hard-sphere polymer into and out of an ellipsoidal 
cavity. Monte Carlo simulations were employed to calculate translocation free energy 
functions, and these functions were used together with the Fokker-Planck equation
to predict the translocation times. We considered the case of ejection from the
cavity for an athermal system, as well as the case of polymer insertion driven
either by a force acting in the pore or by monomer attraction to the walls of the cavity.
We studied the effects on the free energy and predicted dynamics of varying
all of the relevant system parameters, with special attention to the effect
of the cavity anisometry.

The free energy functions calculated for the athermal system with spherical cavities 
are consistent with those calculated previously by analytical methods\cite{kong2004polymer}
and by simulation.\cite{rasmussen2012translocation,polson2013simulation} We find
that the variation of $F$ with packing fraction is approximately consistent with
a scaling theory prediction that treats the polymer to be in the semi-dilute regime 
inside the cavity. Deviations from the predicted scaling exponents likely arise
from finite-size effects. Increasing the anisometry of the cavity at fixed volume
increases the free energy cost of insertion. The effect is approximately symmetric
with respect to prolate or oblate distortions and diminishes with increasing
cavity packing fraction. This increase in the free energy leads to a prediction
of faster polymer ejection for ellipsoidal cavities compared to spherical cavities.
This is consistent with the ejection rates measured in a previous Langevin dynamics
simulation study.\cite{ali2006polymer} Application of a driving force on monomers in the
pore leads to a local free energy minimum for an intermediate range of force magnitudes,
and the corresponding free energy barrier slows the insertion. Increasing the
shape anisometry of the cavity causes faster polymer insertion at high packing
fractions and moderate driving force, but has negligible effect for larger 
cavities or high driving force.  This result is inconsistent with 
Ref.~\onlinecite{ali2006polymer}, most likely because of out-of-equilibrium
conformational behaviour that is not accounted for in the FP approach. Finally, 
the translocation rate for insertion driven by monomer attraction to the cavity 
walls is much more sensitive to cavity anisometry than for the case of insertion 
driven a force in the pore.

It is intriguing that the predicted qualitative effects of the cavity shape on translocation 
rate are in agreement with observations from dynamics simulations for ejection but not
insertion. As noted, the latter discrepancy is due to nonequilibrium behaviour, most 
likely associated with the encapsulated part of the polymer, which provides the
dominant contribution to the free energy function. The internal dynamics of this
portion likely becomes very sluggish at high densities. Increasing the cavity
anisometry at fixed volume narrows the cavity in at least one dimension, leading
perhaps to slower conformational rearrangement and an increase in nonequilibrium
effects.  However, it seems likely that the ejection process would also exhibit this
behaviour as well for the same simulation model. Evidently, there is a hysteresis
effect at play here, which was also noted in Ref.~\onlinecite{ali2006polymer}.
In future work, we will examine this issue further and quantify the deviation
from equilibrium behaviour by carrying dynamics simulations for variable pore
friction. At high pore friction, the polymer ejection dynamics are perfectly 
consistent with FP predictions,\cite{polson2013polymer} and it is likely to be the 
case for insertion as well. This work should contribute to understanding the
limits of the validity of employing free energy arguments to understand DNA
ejection and packaging in viruses, as well as other systems, where nonequilibrium
effects are appreciable. 

\begin{acknowledgments}
This work was supported by the Natural Sciences and Engineering Research Council of 
Canada (NSERC).  We are grateful to the Atlantic Computational Excellence Network 
(ACEnet) for use of their computational resources.
\end{acknowledgments}


%

\end{document}